# A Kernel Ridge Regression Combining Nonlinear ROMs for Accurate Flow Field Reconstruction with Discontinuities

Weiji Wang, Chunlin Gong, Xuyi Jia, Chunna Li[*]

Shaanxi Aerospace Flight Vehicle Design Key Laboratory, School of Astronautics, Northwestern Polytechnical University, Xi'an 710072, People's Republic of China

[*]Email address: chunnali@nwpu.edu.cn

**Abstract**

Nonlinear reduced-order models (ROMs), represented by manifold learning (ML), can effectively improve the modeling accuracy of nonlinear flow fields with discontinuities. However, the inverse mapping from low-dimensional manifold coordinates to high-dimensional flow fields often introduces considerable reconstruction errors, leading to inaccuracy in the locations of discontinuities. To address this challenge, a novel reconstruction method is proposed to enhance the accuracy of reconstructing flow fields with discontinuities. The method employs kernel ridge regression (KRR) to construct a set of nonlinear modes rich in discontinuity information, sequentially these modes are nonlinearly combined with manifold coordinates to achieve accurate flow field reconstruction. The proposed reconstruction method is validated to reconstruct the transonic flow fields over RAE2822 airfoil. Comparison results demonstrate that the method achieves superior reconstruction accuracy compared to existing approaches, especially in reconstructing flow fields' discontinuous regions and precisely capturing discontinuities. This work provides an effective and highly interpretable solution for improving the accuracy of nonlinear ROMs in discontinuous flow fields modeling.

**Key words:** flow field reconstruction, discontinuities, kernel ridge regression, manifold learning

## 1. Introduction

Accurate predicting of flow fields, particularly those containing discontinuities like shocks, is crucial for aerodynamic design. These discontinuities significantly impact the overall flow behavior and performance, demanding high-fidelity simulation tools. Although traditional computational fluid dynamics (CFD) methods have developed rapidly in the past decades, its high computational expense limits their applications in design optimization processes especially for engineering problems. Therefore, the development of reduced-order models (ROMs) has been motivated in recent years. ROMs aims to capture the essential characteristics of the flow fields using a significantly reduced number of dimensions compared to full-order CFD simulations, which leads to substantial improvements in computational efficiency.

Traditional ROMs operate by projecting the high-dimensional flow fields onto a lower dimensional subspace spanned by a set of linear modes. These modes are typically obtained by proper orthogonal decomposition (POD)[1] or dynamic mode decomposition (DMD)[2]. In existing studies, POD and DMD have been widely applied to flow field modeling [3–8], prediction [9–14] and analysis [15–20]. However, since the sharp gradients and non-smooth nature of those discontinuities are poorly approximated by the smooth modes employed by POD and DMD, these POD-based or DMD-based methods often struggle to accurately reconstruct flow fields with discontinuities. At the same time, the linear characteristics of POD and DMD also presents a challenge in capturing the nonlinear features of those discontinuities. These leads to significant errors in the reconstructed flow fields, particularly in the regions near the discontinuities. Although methods such as local partitioning[21], weighted POD[22,23] and combinatorial method[24] have been proposed to address nonlinear flow fields prediction and analysis, the linear feature of these methods limits their further application in modeling flow fields with discontinuities.

To address this issue, researchers begin exploring nonlinear ROMs, which mainly include manifold learning (ML). ML assumes that high-dimensional data is distributed on a low-dimensional manifold, which can capture the intrinsic features and relationships between the data. The pioneer works on ML include isometric mapping (ISOMAP)[25] and local linear embedding (LLE)[26] which are both published in 2000. ISOMAP originates from the classical multiple dimensional scaling (MDS)[27]. It replaces the Euclidean distance used in MDS with the geodesic distance, allowing it to achieve a low-dimensional manifold of high-dimensional nonlinear data while simultaneously preserving both local linear structures and global nonlinear structures. Unlike ISOMAP, which preserves the distances between nearby samples, LLE extracts the manifold structure by maintaining the topological relationships among samples within a neighborhood. LLE assumes that each sample in the high-dimensional space can be reconstructed by a linear combination of its neighboring samples, and such a linear relationship can be preserved in the low-dimensional manifold space. These two ML methods have already been applied in the dimensionality reduction and analysis of nonlinear flow fields. Tauro[28] employed ISOMAP to identify flow patterns while Marra[29] describes the relationship between manifold coordinates and physical parameters. Similar work by Farzamnik[30] finds that there is a correlation between the manifold coordinates and dimensionless force coefficients. Ehlert[31] adopts LLE to identify the reduced-order representations of cylinder wake transients. In the modeling of nonlinear flow, Franz[32] firstly applies ISOMAP to predict the transonic steady flow fields. The prediction results for the airfoil and wing indicate that the pressure distribution obtained using nonlinear ROMs is more accurate, especially near the shock, compared to linear ROMs. This work demonstrates the potential of the ML-based ROMs in modeling flow fields with discontinuities. Therefore, more flow fields modeling methods based on ML are developed. For example, Decker[33] combines ML and manifold alignment to build a multi-fidelity nonlinear ROM. Zheng[34] adopts constrained optimization genetic algorithm (COGA), using the distances between sample and its neighbors as constraints to optimize the interpolated manifold coordinates, aiming to achieve more accurate prediction results with small sample sizes. Li[35] uses ISOMAP to extract global features and LLE to extract local features of the flow fields, achieving accurate prediction of transonic and supersonic flow fields.

Unlike POD and DMD, MLs do not provide an explicit mapping relationship between the low-dimensional manifold coordinates and the high-dimensional flow fields. Therefore, when applying ML methods for flow fields modeling, specific inverse mapping algorithms are required. The previously-mentioned modeling cases for flow fields with discontinuities mostly use the non-parametric back-mapping (NPBM) algorithm proposed by Franz[32] as the inverse mapping algorithm. This algorithm draws inspiration from the idea of LLE, assuming that the high-dimensional flow fields can be reconstructed through a linear combination of its surrounding neighbors. However, when there are significant differences in the intensity and position of discontinuities between neighbors, this method tends to result in large deviations in the discontinuity positions and significant errors in the reconstructed flow fields. Currently, few studies have focused on reconstruction methods between the low-dimensional manifold coordinates of flow fields with discontinuities and the original high-dimensional flow fields. How to achieve accurate reconstruction of the discontinuities' intensity and location is an issue requiring attention.

To address above problem, this paper proposes a novel reconstruction method for flow fields with discontinuities. The method constructs a set of nonlinear modes using Kernel Ridge Regression (KRR) based on the low-dimensional manifold coordinates obtained from nonlinear ROMs and the original flow fields. Then, the flow field reconstruction is achieved by the nonlinear combination of low-dimensional manifold coordinates and these modes. Compared to POD and NPBM, the proposed reconstruction method can more accurately reconstruct the flow field near discontinuities and precisely predict the location of the discontinuities.

The paper is organized as follows: in section 2, a detailed description of the developed reconstruction method is provided; the following section includes method validation on transonic flow fields around RAE2822 airfoil, key parameters analysis and a comparison with existing POD and NPBM methods; the conclusion and future directions are giving in the last section. Appendix A describes the implementation procedures of ISOMAP and LLE, and Appendix B shows the accuracy verification of CFD solver.

## 2. Methodology

In this work, an accurate reconstruction methodology for flow fields with discontinuities is developed. The proposed approach consists of two steps. First, dimensionality reduction of flow fields with discontinuities is performed by ML. This aims to find the hidden low-dimensional nonlinear manifold of the original high-dimensional flow fields. Second, the original flow field is accurately reconstructed by previously obtained manifold coordinates. This reconstruction procedure is achieved by combining a set of nonlinear modes obtained by kernel ridge regression (KRR). Figure 1 shows the main stages of our flow field reconstruction procedure, and the details are introduced as follows.

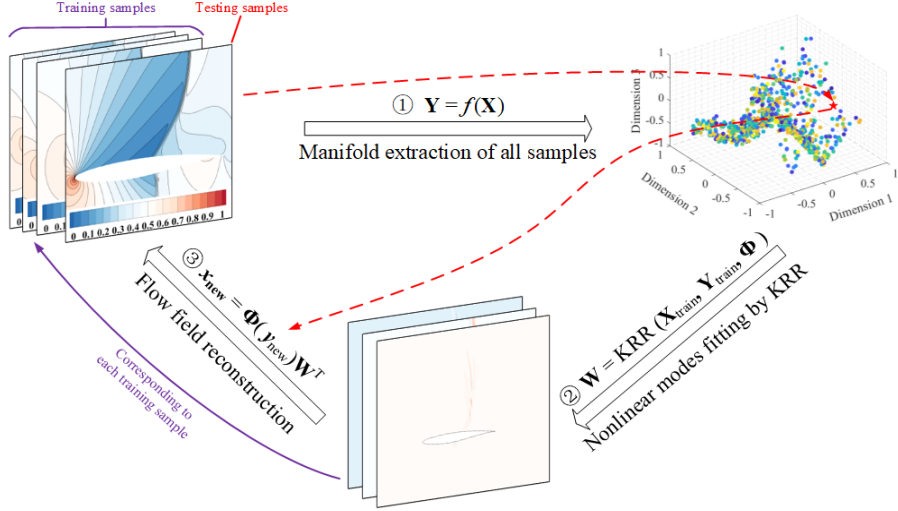

Figure 1 Procedure of the proposed flow field reconstruction method

### 2.1. Dimensionality reduction based on ML

Since flow fields with discontinuities exhibits strong nonlinearity, traditional linear dimensionality reduction techniques are not applicable. Therefore, it is necessary to use nonlinear methods, among which ML is one of the most representative approaches.

Consider that $n$ flow fields samples are obtained by CFD. Each sample is vectorized to an $m$-dimensional vector $x_i$ (subscript $i$ indicates the $i$th sample) and all of them form a samples matrix $\mathbf{X} \in \mathbf{R}^{n \times m}$. ML assumes that $\mathbf{X}$ is embedded on a $p$-dimensional manifold and each axis of the manifold captures the main characteristics of the flow fields. Generally, $p$ is much smaller than $m$ so if we can build a mapping $f(\cdot)$ between the original high-dimensional flow fields and low-dimensional manifold, the dimensionality reduction of the flow fields can be achieved:

$$\mathbf{Y} = f(\mathbf{X}) \tag{1}$$

where $\mathbf{Y} \in \mathbf{R}^{n \times p}$ is the low-dimensional representation of $\mathbf{X}$.

According to the methods of constructing $f(\cdot)$, ML can be divided into various types. LLE and ISOMAP are two of the most commonly used ML methods. The main processes of LLE and ISOMAP are specifically described in Appendix A. The comparison of different ML methods is presented in section 3.2.

### 2.2. Flow field reconstruction via KRR

Since the low-dimensional representation $\mathbf{Y}$ are obtained by ML, we need to build an inverse mapping model to reconstruct the original flow fields $\mathbf{X}$. Through this model, the low-dimensional representation of any flow field $\mathbf{Y}_{new}$ can be reconstructed.

Drawing inspiration from the idea of POD, we aim to find a modes matrix $\mathbf{W}$ $\mathbf{W} \in \mathbf{R}^{p \times m}$ that satisfies

$$\mathbf{X} \approx \mathbf{X}' = \mathbf{Y}\mathbf{W}^{\mathrm{T}} \tag{2}$$

where $\mathbf{X}' \in \mathbf{R}^{n \times m}$ is the reconstructed flow fields matrix.

The above reconstruction method is a linear method, where $\mathbf{W}$ represents a linear modes matrix. However, in order to achieve accurate reconstruction in nonlinear regions considering discontinuities, it is desirable to introduce a nonlinear modes matrix $\mathbf{W}_{\text{nonlinear}}$. Thus, we map $\mathbf{Y}$ from the manifold space to the kernel inner product space using a kernel function $\Phi(\cdot)$, and construct a nonlinear modes matrix in this space to achieve the nonlinear reconstruction

$$\mathbf{X} \approx \mathbf{X}' = \Phi(\mathbf{Y}, \mathbf{Y})\mathbf{W}_{\text{nonlinear}}^{\text{T}} = \mathbf{Z}\mathbf{W}_{\text{nonlinear}}^{\text{T}} \tag{3}$$

where $\mathbf{Z}$ is the kernel matrix, which represents the nonlinear similarity between training samples.

We use ridge regression to solve $\mathbf{W}_{\text{nonlinear}}$ by minimize the following target function

$$\left\| \mathbf{X} - \mathbf{Z}\mathbf{W}_{\text{nonlinear}}^{\text{T}} \right\|_{\text{F}}^{2} + \lambda \left\| \mathbf{W}_{\text{nonlinear}} \right\|_{\text{F}}^{2} \tag{4}$$

where $\left\| \cdot \right\|_{\text{F}}$ represents the Frobenius norm; $\lambda$ is the L2 regularization coefficient. Formula (4) consists of two parts: the first indicates that the difference between the original flow fields and the reconstructed flow fields is sufficiently small; the second prevents overfitting. The influence of $\lambda$ is further discussed in section 3.4.

Let formula (4) equals 0 and the expression for $\mathbf{W}_{\text{nonlinear}}$ is

$$\mathbf{W}_{\text{nonlinear}} = \mathbf{X}^{\text{T}}\mathbf{Z}(\mathbf{Z}^{\text{T}}\mathbf{Z} + \lambda \mathbf{I})^{+} \tag{5}$$

where $\mathbf{I} \in \mathbf{R}^{n \times n}$ is an identity matrix and $(\cdot)^{+}$ represents pseudo-inverse. $\mathbf{W}_{\text{nonlinear}}$ is an $m \times n$ matrix, where each column represents a nonlinear mode. The physical significance of each mode is explained in section 3.5.

Assuming we obtain the low-dimensional representations of $k$ new flow fields in manifold space, which form a matrix $\mathbf{Y}_{\text{new}} \in \mathbf{R}^{k \times p}$. The reconstructed flow fields matrix $\mathbf{X}'_{\text{new}} \in \mathbf{R}^{k \times p}$ is

$$\mathbf{X}'_{\text{new}} = \mathbf{Z}_{\text{new}}\mathbf{W}_{\text{nonlinear}}^{\text{T}} = \Phi(\mathbf{Y}_{\text{new}}, \mathbf{Y})\mathbf{W}_{\text{nonlinear}}^{\text{T}} \tag{6}$$

where $\mathbf{Z}_{\text{new}}$ is the kernel matrix between new samples and training samples.

### 2.3. Kernel methods

During the construction of $\mathbf{W}_{\text{nonlinear}}$ in the above section, we use a kernel function to project manifold space into kernel inner product space. In this section, we talk about the basic idea of kernel methods.

The mathematical foundation of kernel methods lies in the concept of mapping data into a higher-dimensional space $\mathcal{H}$ and leveraging inner products in that space to solve linear problems that correspond to nonlinear problems in the original space $\mathcal{X}$ by mapping function $\phi$

$$\phi : \mathcal{X} \rightarrow \mathcal{H} \tag{7}$$

Since most of $\phi$ cannot be explicitly computed, kernel methods use a kernel function $\mathrm{K}(\cdot)$ to compute the inner product of two samples directly in $\mathcal{H}$

$$\mathrm{K}(\boldsymbol{x}, \boldsymbol{y}) = \langle \phi(\boldsymbol{x}), \phi(\boldsymbol{y}) \rangle \tag{8}$$

the value of formula (8) shows the similarity between $\boldsymbol{x}$ and $\boldsymbol{y}$.

For two data matrices $\mathbf{A} \in \mathbf{R}^{r \times d}$ and $\mathbf{B} \in \mathbf{R}^{q \times d}$, which contain $r$ and $q$ data points respectively, with each data point having a dimension of $d$, the similarity between the data in $\mathbf{A}$ and $\mathbf{B}$ can be represented using a kernel matrix $\mathbf{Z}$

$$\mathbf{Z} = \Phi(\mathbf{A}, \mathbf{B}) \tag{9}$$

$$\mathbf{Z}_{ij} = \mathrm{K}(\boldsymbol{a}_i, \boldsymbol{b}_j) \tag{10}$$

where $\Phi(\cdot,\cdot)$ represents the kernel function of matrices; $\mathbf{Z}_{ij}$ is the element in the $i$th row and $j$th column of the kernel matrix $\mathbf{Z}$. $\boldsymbol{a}_i \in \mathrm{R}^d$ is the $i$th sample in $\mathbf{A}$; $\boldsymbol{b}_j \in \mathrm{R}^d$ is the $j$th sample in $\mathbf{B}$.

In this work, we employ 4 commonly-used kernel functions of which the first three are nonlinear kernel functions, while the last one is a linear kernel function. The mathematical expressions of each kernel function and the visualizations are shown in Table 1 and Figure 2. The comparison of the effect by different kernels in flow field reconstruction is presented in section 3.2, and the impact of $\gamma$ on the reconstruction accuracy is discussed in section 3.3.

Table 1 Different kernel functions used in this paper

| Kernel type | Expression |
| --- | --- |
| Polynomial kernel | $\mathrm{K}(\boldsymbol{a}_i, \boldsymbol{b}_j) = (\gamma \boldsymbol{a}_i^{\mathrm{T}} \boldsymbol{b}_j)^l$ |
| RBF kernel | $\mathrm{K}(\boldsymbol{a}_i, \boldsymbol{b}_j) = \exp(-\gamma \|\boldsymbol{a}_i - \boldsymbol{b}_j\|_{\mathrm{F}}^2)$ |
| Sigmoid kernel | $\mathrm{K}(\boldsymbol{a}_i, \boldsymbol{b}_j) = \tanh(\gamma \boldsymbol{a}_i^{\mathrm{T}} \boldsymbol{b}_j)$ |
| Linear kernel | $\mathrm{K}(\boldsymbol{a}_i, \boldsymbol{b}_j) = \gamma \boldsymbol{a}_i^{\mathrm{T}} \boldsymbol{b}_j$ |

*$\gamma$ is the scale factor and $l$ is the degree of the polynomial

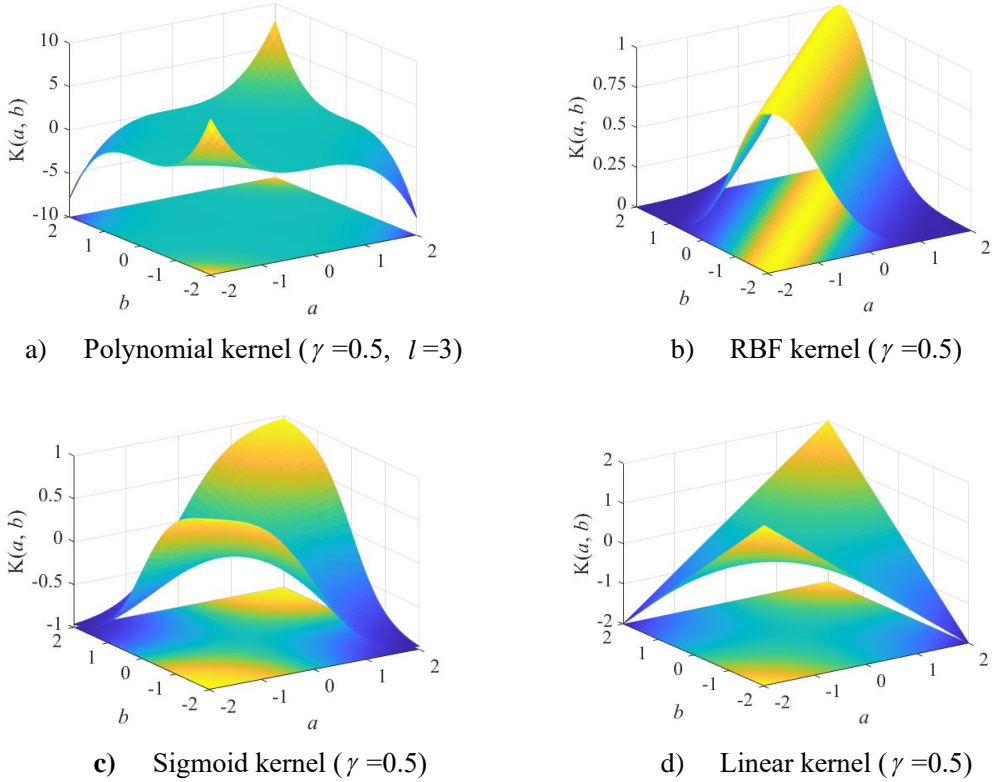

a) Polynomial kernel ($\gamma = 0.5$, $l = 3$)  b) RBF kernel ($\gamma = 0.5$)

c) Sigmoid kernel ($\gamma = 0.5$)  d) Linear kernel ($\gamma = 0.5$)

Figure 2 Visualizations of different kernel functions

## 3. Method Validation and Parameter Analysis

This section presents the validation case and discusses the performance of proposed flow field reconstruction method. The first subsection describes the generation of the dataset for the transonic flow fields around RAE2822, and gives the accuracy evaluation method which provides quantitative criteria for subsequent parameter analysis and model comparison. The following two subsections implement the reconstruction of the transonic flow field and analyze the impact of the key model parameters on reconstruction accuracy. The last subsection compares the proposed method with the existing flow field reconstruction methods, and explains the physical significance of the constructed nonlinear modes obtained by KRR. All the contours shown in this section are the normalized pressure fields. The normalization is implemented according to

$$p_{\text{Norm}}^{(i)} = \frac{p^{(i)} - \min(p_{\text{test}}^{(i)}) \cdot \mathbf{1}}{\max(p_{\text{test}}^{(i)}) - \min(p_{\text{test}}^{(i)})} \quad (11)$$

where $p_{\text{Norm}}$ is the normalized flow field vector; $p_{\text{test}}$ is the original flow field vector of a testing sample; $\mathbf{1}$ is an all-ones vector of the same length as $p_{\text{test}}$; superscript ($i$) represents the $i$th testing sample.

### 3.1. Dataset generation and accuracy evaluation criterion

The reconstruction of the transonic flow fields around the RAE2822 airfoil under $M_\infty \in [0.734, 0.85]$ and $\alpha \in [2°, 6°]$ is studied. Within this range of flight conditions, the position and intensity of shocks in the flow field change drastically. The Latin hypercube sampling (LHS) is employed to select 900 samples. Since this work only focuses on the process of flow field reconstruction and does not involve flow field prediction, all 900 samples are used for dimensionality reduction, while 50 of them are randomly selected as testing samples for flow field reconstruction. The distribution of all the 900 samples in the parameter space is shown in Figure 3.

We adopt a density-based compressible Reynolds-average Navier-Stokes (RANS) solver with SST $k$-$\omega$ turbulence model for CFD simulations. The computational mesh is shown in Figure 4. The accuracy verification of CFD solver is presented in Appendix B.

The accuracy of the flow field reconstruction is qualified by measuring mean relative reconstruction error $E_{\text{mean}}$ and maximum relative reconstruction error $E_{\text{max}}$ on the testing samples. The definition of $E_{\text{mean}}$ and $E_{\text{max}}$ are given as follows

$$E_{\text{mean}} = \frac{\sum_{k=1}^{n}\sum_{i=1}^{m} e_i^{(k)}}{nm} \quad (12)$$

$$E_{\text{max}} = \frac{\sum_{k=1}^{n} \max_{i \in [1,m]}(e_i^{(k)})}{n} \quad (13)$$

$$e_i^{(k)} = \left| \frac{x_i^{\text{real}} - x_i^{\text{recons}}}{x_i^{\text{real}}} \right| \tag{14}$$

where $n$ is the number of testing samples; $m$ is the number of mesh cells; $e_i^{(k)}$ represents relative reconstruction error on the $i$th cell of the $k$th testing sample; $x$ is the flow field quantity which could be pressure, temperature, density and so on. In this paper, $x$ refers to be pressure. The superscripts "real" and "recons" represent the original flow field data and the reconstructed flow field data, respectively.

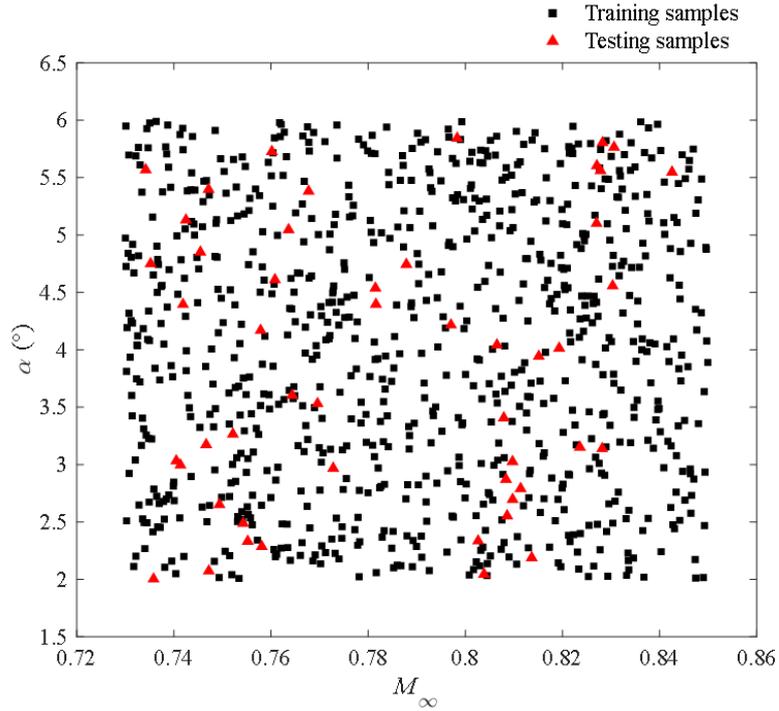

Figure 3 Training and testing samples in the $M_\infty$-$\alpha$ parameter space

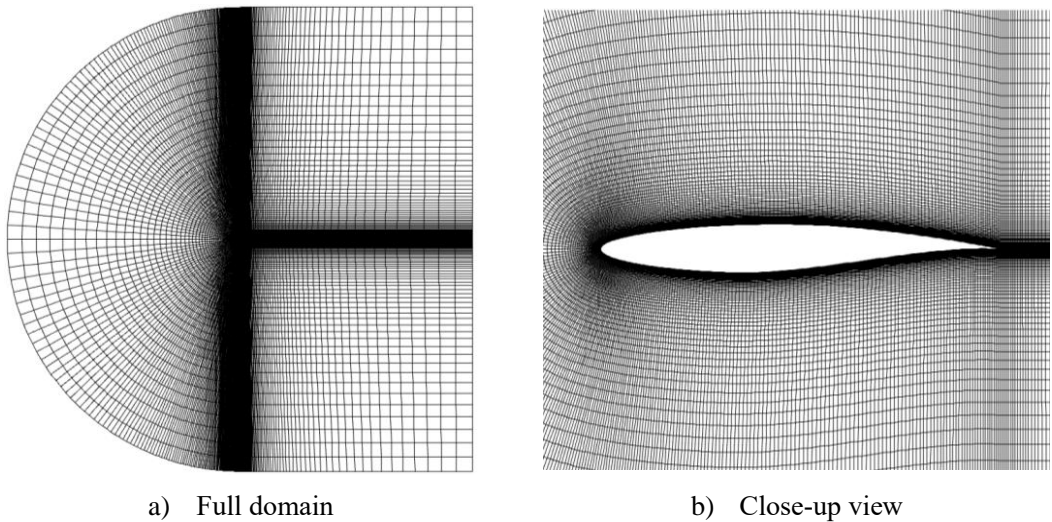

a) Full domain  b) Close-up view

Figure 4 The computational mesh for the RAE2822 airfoil (82,844 cells in total)

## 3.2. Flow field reconstruction

We firstly perform dimensionality reduction on the dataset using the LLE and ISOMAP with 15 neighbors respectively. Then we reconstruct the flow fields of the 50 testing samples using the proposed reconstruction method. For both cases, kernel function is set to cubic polynomial kernel and $\gamma=1$. Calculate $E_{mean}$ and $E_{max}$ under different manifold dimensions $d$ and the results are shown in Figure 5. When $d$ is greater than 15, $E_{mean}$ of the ISOMAP starts to gradually increase; when $d$ is between 10 and 20, $E_{max}$ of the ISOMAP exhibits an abnormal peak. However, compared to the ISOMAP, both $E_{mean}$ and $E_{max}$ are much smaller when using the LLE. This indicates that, for the proposed flow field reconstruction method, it is more suitable to use the LLE to perform dimensionality reduction of flow fields.

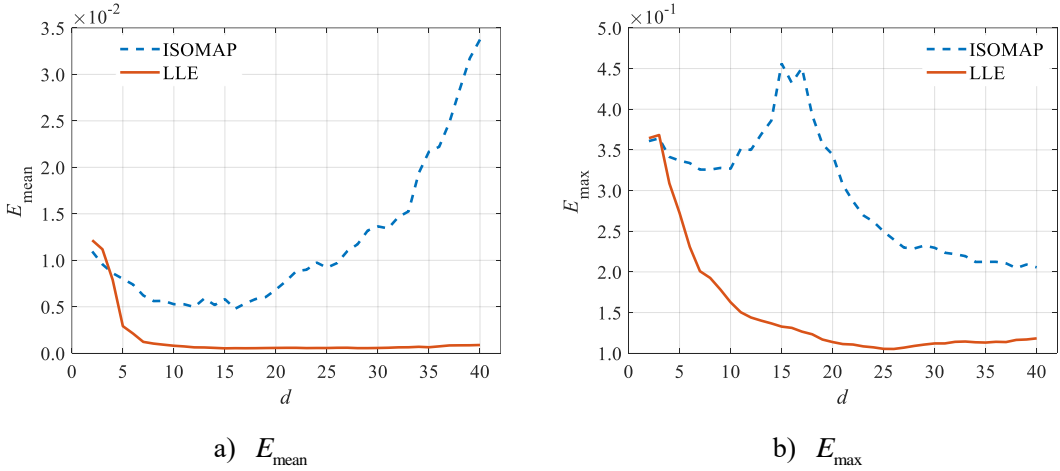

a) $E_{mean}$  b) $E_{max}$

Figure 5 Relative reconstruction error through different ML methods

Then, the impact of kernel function types on the reconstruction accuracy is studied. We compare the relative reconstruction error of 4 different kernel functions in Table. 1, where the degree of the polynomial kernel is taken sequentially from 2$^{nd}$ to 5$^{th}$ order, as shown in Figure 6. $\gamma$ of all the kernels are set to 1, then the LLE is used to perform dimensionality reduction. Linear kernel shows the poorest performance compared to other 4 nonlinear kernels on $E_{max}$. When $d$ is less than 35, $E_{mean}$ of the linear kernel is larger than other nonlinear kernels, except RBF and sigmoid kernel. This result is consistent with the nonlinear characteristics of the flow fields with discontinuities. However, the linear kernel can still achieve an accurate reconstruction result globally when $d$ is large, as shown in Figure 7, since the nonlinear characteristics in the regions away from discontinuities are relatively weak.

For the nonlinear RBF and sigmoid kernels, $E_{mean}$ starts to increase significantly after $d$ exceeds 10. This is because, as the dimension of the manifold space increases, the distribution of samples becomes sparse and uniform. This causes the RBF and sigmoid kernels struggle in obtaining more global information, which is reflected in the reconstruction error as an increase in $E_{mean}$. However, these two types of kernels can still obtain sufficient local information, causing a continuously decrease of $E_{max}$.

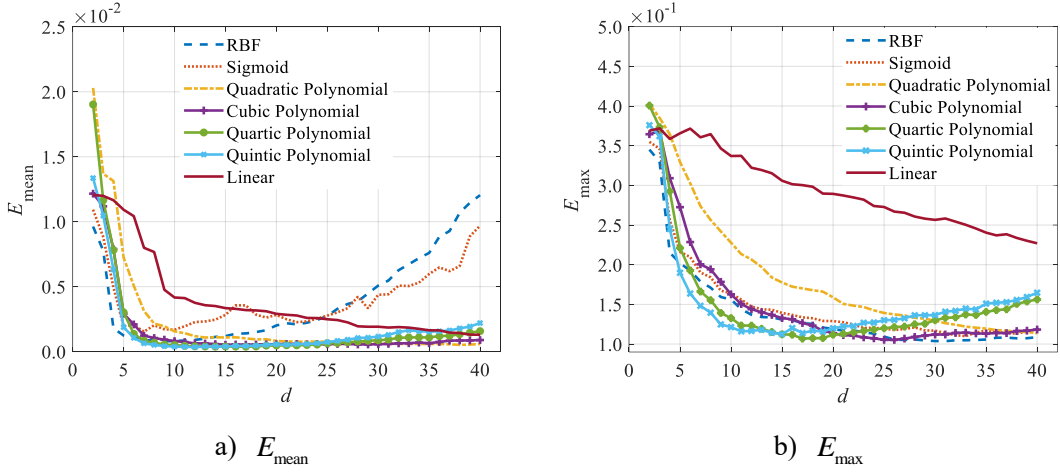

a) $E_{mean}$

b) $E_{max}$

Figure 6 Relative reconstruction error through different kernels

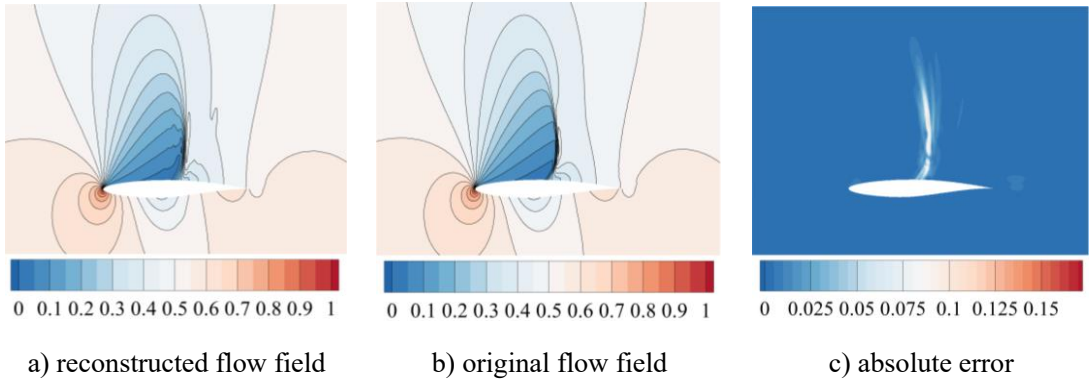

a) reconstructed flow field

b) original flow field

c) absolute error

Figure 7 Flow field reconstruction result through linear kernel with $d$=35 at $M_\infty = 0.758$ and $\alpha = 4.169°$

Compared to the RBF and sigmoid kernels, using a polynomial kernel can achieve a relatively smaller $E_{mean}$ under the same level of $E_{max}$. The further comparison of the polynomial kernels with different degrees is shown in Figure 8, with the y-axis using a logarithm to base 10. For quartic and quintic polynomial kernels, $E_{mean}$ and $E_{max}$ begin to increase when $d$ exceed around 16, which is caused by the amplification of data noise and overfitting. Using a low-degree polynomial kernel can alleviate this problem. In our case, we hope to reconstruct the flow fields accurately by adopting a smaller $d$. Therefore, we choose the quartic polynomial kernel to reconstruct the flow fields at $d$=17, which can simultaneously meet the requirements of $E_{mean}$ and $E_{max}$. The reconstructed flow field at $M_\infty = 0.758$ and $\alpha = 4.169°$ is shown in Figure 9. It indicates that the largest reconstruction error in the flow fields occurs near the discontinuities, which is the same as Figure 7. It illustrates that the proposed nonlinear flow field reconstruction method can effectively improve the reconstruction accuracy in the regions with discontinuities.

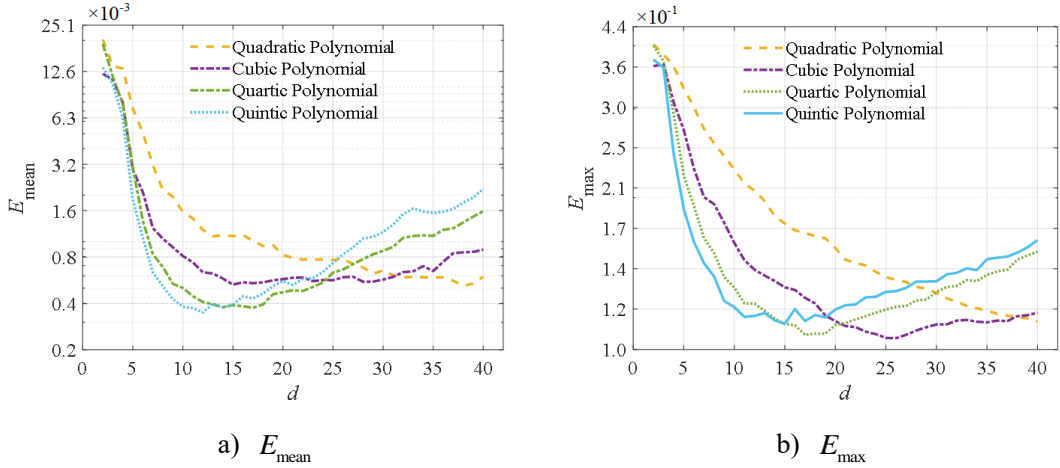

a) $E_{mean}$

b) $E_{max}$

Figure 8 Relative reconstruction error under different degrees of polynomial kernel (base-10 logarithmic coordinate)

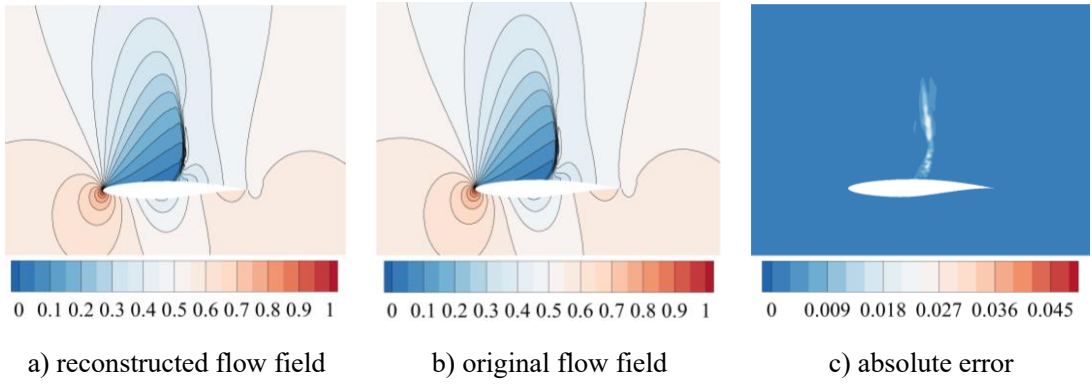

a) reconstructed flow field    b) original flow field    c) absolute error

Figure 9 Flow field reconstruction result through quartic kernel with $d$=17 at $M_\infty = 0.758$ and $\alpha = 4.169°$

### 3.3. Parametric analysis of $\lambda$ and $\gamma$

In our proposed method, there are two key hyperparameters $\lambda$ and $\gamma$, which affect the reconstruction accuracy. In this section, we provide recommendations for their selection through parametric analysis.

The L2 regularization coefficient $\lambda$ can control the influence strength of the regularization term when solving the nonlinear mode matrix $\mathbf{W}_{nonlinear}$, thereby balancing the model's fitting capability and complexity. The relative reconstruction error on the testing samples under different $\lambda$ is shown in Figure 10. When $\lambda$ is larger, both $E_{mean}$ and $E_{max}$ decrease as $\lambda$ decreases. $E_{mean}$ reaches minimum when $\lambda$ is around $10^{-5}$. When $\lambda<10^{-5}$, as $\lambda$ continues to decrease, $E_{mean}$ begins to grow, while $E_{max}$, although continuing to decrease, does so with a less noticeable decline. This is because when $\lambda$ is too small, the regularization term loses its effect, and is overfitted under the influence of noise.

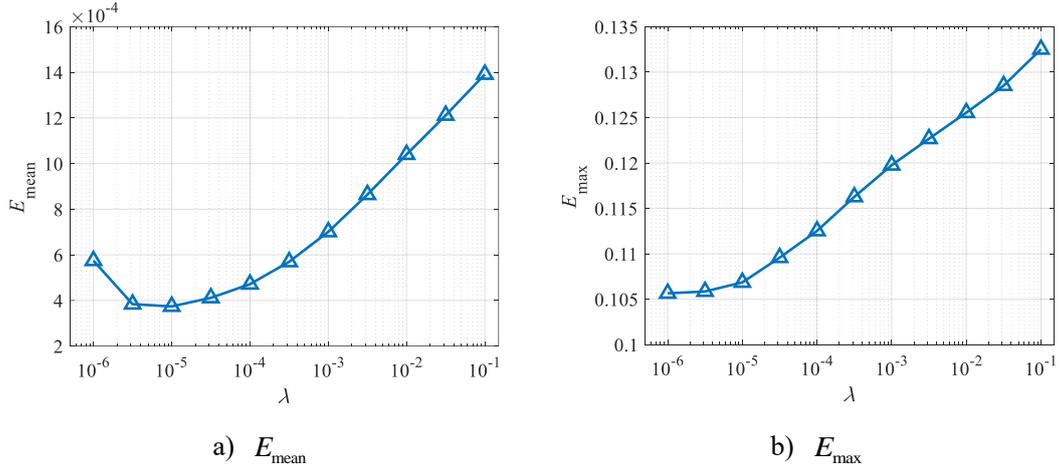

a) $E_{mean}$      b) $E_{max}$

Figure 10 Relative reconstruction error under different $\lambda$

The scale factor $\gamma$ determines the extent to which the inner product of samples affects the polynomial kernel. When $\gamma$ is larger, the inner product between samples is amplified, and the kernel function becomes more sensitive to local samples with high similarity. The variations in $E_{mean}$ and $E_{max}$ for different $\gamma$ are shown in Figure 11. When $\gamma > 1.1$, $E_{mean}$ is basically flat (the difference between the maximum and minimum values is less than 0.00025). While $E_{max}$ slightly increases when $\gamma > 1.4$. It is because the number of locally high-similarity samples is limited, and excessive sensitivity to these samples prevents the model from obtaining sufficient discontinuity's information to reconstruct regions with discontinuities. Overall, it is more rational to set $\gamma = 1.1$.

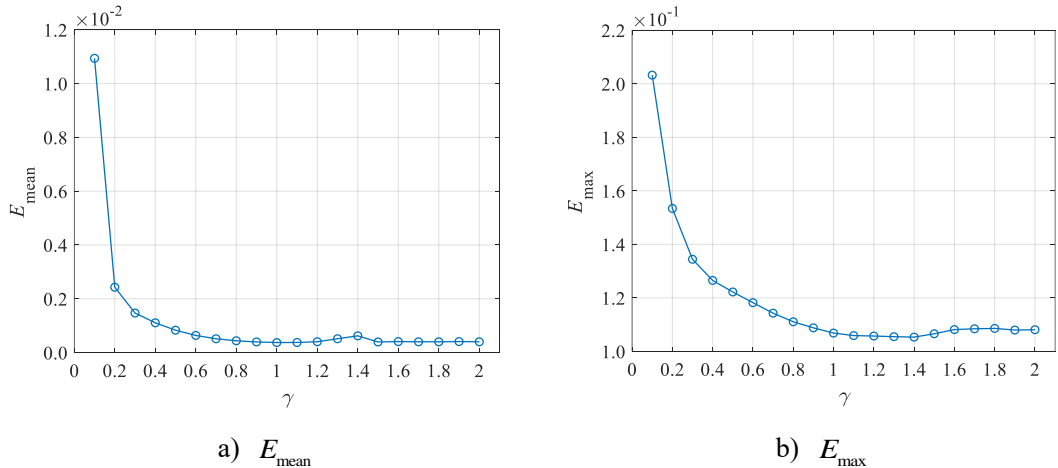

a) $E_{mean}$      b) $E_{max}$

Figure 11 Relative reconstruction error under different $\gamma$

### 3.4. Comparison with existing reconstruction methods

According to the analysis in Sections 3.2 and 3.3, a reconstruction model is established and compared with POD and NPBM methods. When using NPBM, the low-dimensional manifold of the flow fields is obtained using ISOMAP which is the same as Franz[32]. 15 neighbors are adopted to perform dimensionality reduction and flow field reconstruction. For all methods, the high-dimensional flow fields are reduced to 17 dimensions. 5 testing samples from Section 3.1 are selected as reconstruction samples, and the distribution of them in $M_\infty$-$\alpha$ parameter space is shown in Figure 12.

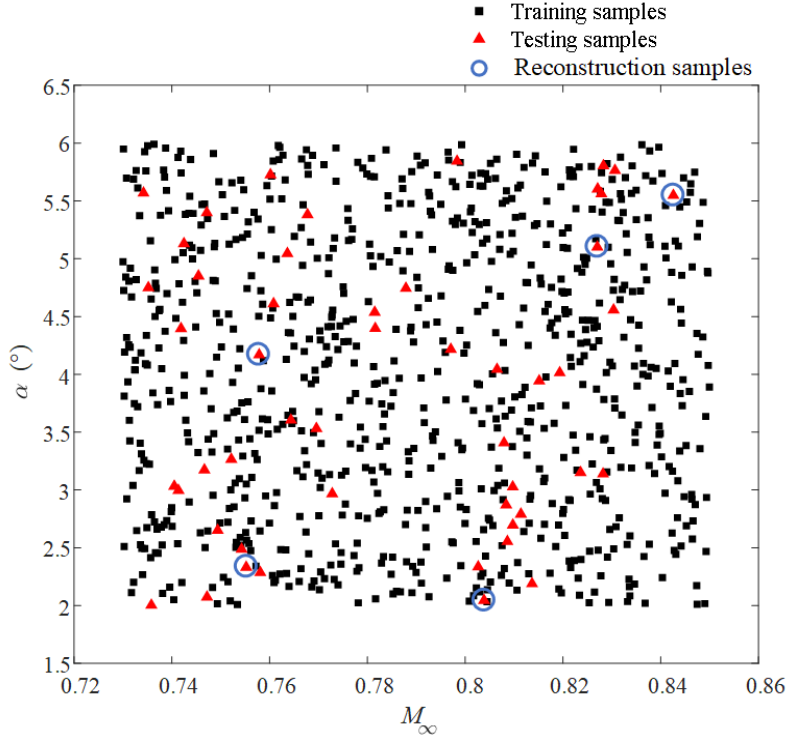

Figure 12 Reconstruction samples in $M_\infty$-$\alpha$ parameter space

The reconstruction results of wall pressure by different methods are illustrated in Figure 13. The reconstructed wall pressure of all 5 reconstruction samples using POD show strong oscillations near the discontinuities. The causes of this phenomenon can be attributed to two aspects. First, the truncation of high-order modes in POD leads to the loss of discontinuous information, making it difficult to reconstruct the details of the discontinuities in the flow fields. Second, the POD modes exhibit globally smooth characteristics, and the linear combination of these finite-order smooth modes is insufficient to accurately capture and reconstruct the local discontinuities. These result in oscillations in the reconstructed flow fields similar to the Gibbs phenomenon observed in signal processing[36].

Compared to POD, both NPBM and KRR demonstrate very high accuracy in reconstructing wall pressure, especially near the discontinuities. This is attributed to the nonlinear ML ROM, which effectively preserves the discontinuous information in the flow fields. However, at higher Mach numbers, as shown in Figure 13 d) and e), NPBM exhibits a significant error in the reconstruction of the shock on the lower surface, while KRR is still able to maintain a high level of reconstruction accuracy.

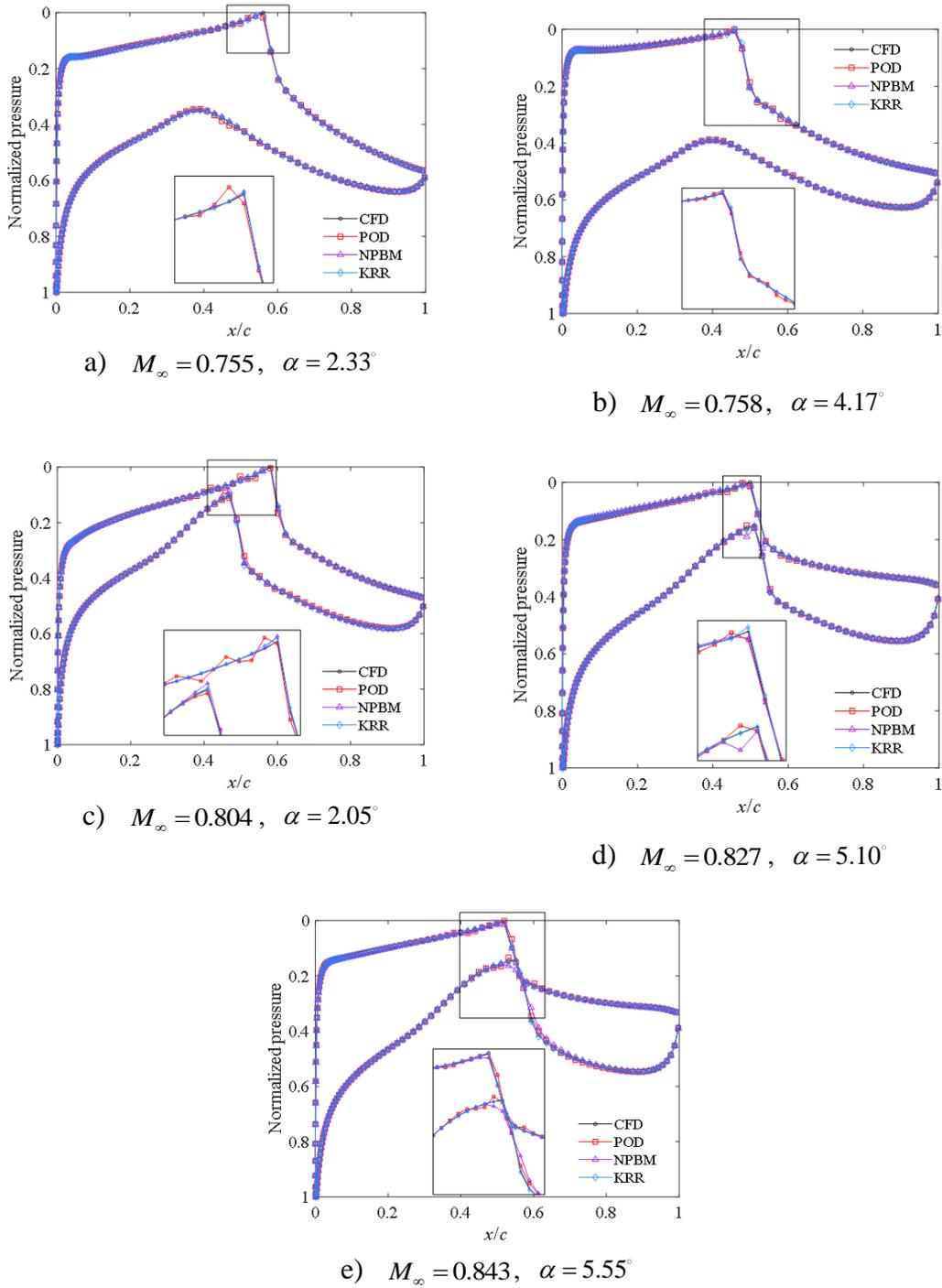

a)  $M_\infty = 0.755$,  $\alpha = 2.33°$

b)  $M_\infty = 0.758$,  $\alpha = 4.17°$

c)  $M_\infty = 0.804$,  $\alpha = 2.05°$

d)  $M_\infty = 0.827$,  $\alpha = 5.10°$

e)  $M_\infty = 0.843$,  $\alpha = 5.55°$

Figure 13 Wall pressure reconstruction results of different reconstruction samples

In the reconstruction of the entire flow field, as shown in Figure 14-Figure 18, the differences between these 3 methods are more significant. The isobars of the flow fields reconstructed by POD still exhibit significant oscillations near the discontinuities and such oscillations become more pronounced with the increase in $M_\infty$ and $\alpha$, as shown in Figure 16, Figure 17 and Figure 18. The increase in $M_\infty$ leads to a dramatic increase in the shock intensity on the upper surface of the airfoil, thereby causing stronger nonlinearity in the flow fields, leading to greater loss of discontinuities' information when applying POD.

NPBM can reconstruct flow fields more accurately compared to POD. This benefits from nonlinear dimensionality reduction where can effectively capture discontinuous' characteristics. However, along with the increase of $M_\infty$, the reconstruction accuracy of NPBM significantly decreases, as shown in Figure 17 and Figure 18. Although nonlinear ROM can indeed capture the discontinuities' information of the flow fields, the NPBM, which relies on a linear combination of neighbors, still struggles to accurately recover discontinuities under high shock intensity.

Among the three methods, KRR achieves the highest reconstruction accuracy across all reconstruction samples according to Figure 14-Figure 18. KRR can accurately reconstruct the flow fields in discontinuous regions thanks to not only nonlinear ROM but also the effect of nonlinear modes. Under the influence of the kernel function, these modes can undergo nonlinear combinations, allowing for further enhancement of the reconstruction accuracy in discontinuous regions. However, for reconstruction sample at $M_\infty = 0.843$ and $\alpha = 5.547°$, KRR exhibits relatively large reconstruction errors in the shock region away from upper surface compared to other 4 reconstruction samples, as shown in Figure 18. At the same time, the isobars in this region also show oscillations similar to those observed in POD. This may be caused by the model's inadequate acquisition of discontinuities' information in this region. At high $M_\infty$ and large $\alpha$, the intensity and position of the shock differ significantly from most training samples. Although the training samples neighboring this reconstruction sample can provide some discontinuities' information in this region, their limited quantity restricts further improvement in reconstruction accuracy. Nevertheless, KRR is still more accurate in capturing the shock's position compared to POD and NPBM.

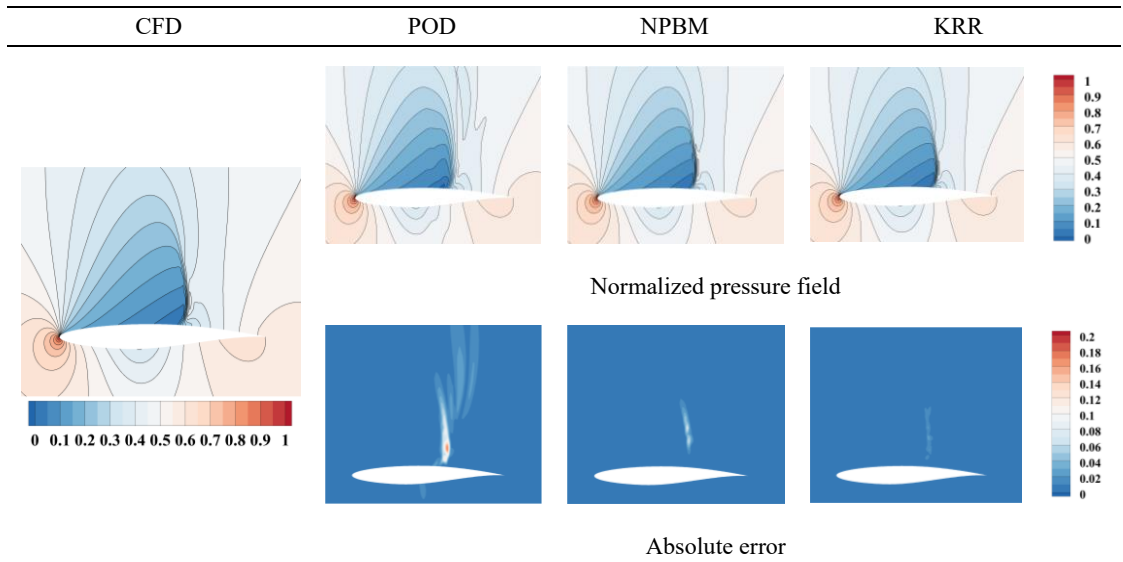

Figure 14 Entire flow field reconstruction results of different methods at $M_\infty = 0.755$, $\alpha = 2.33°$ and comparison with original flow field

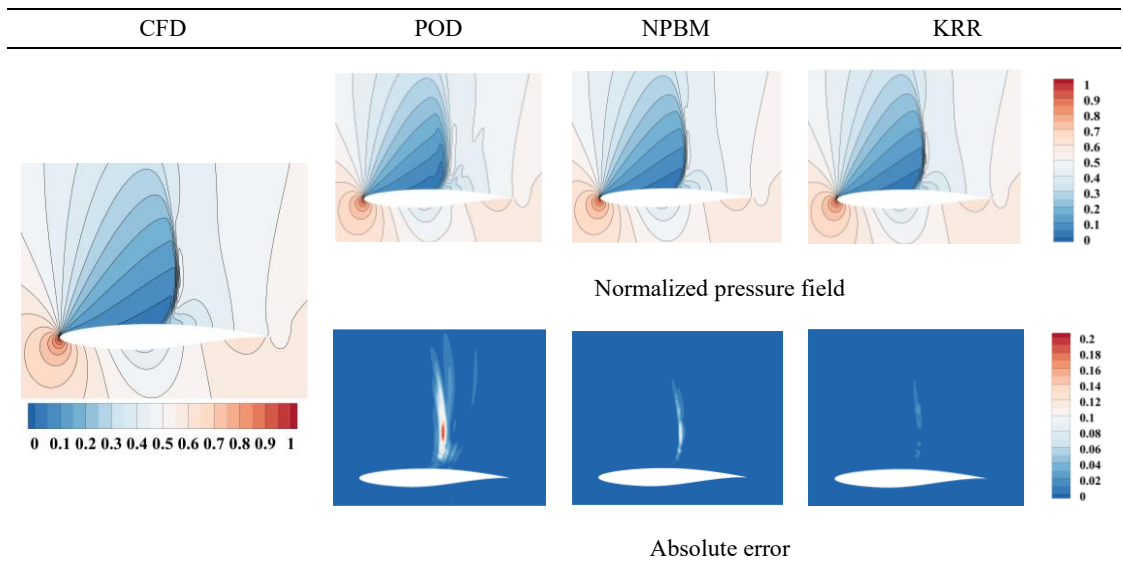

Figure 15 Entire flow field reconstruction results of different methods at $M_\infty = 0.758$, $\alpha = 4.17°$ and comparison with original flow field

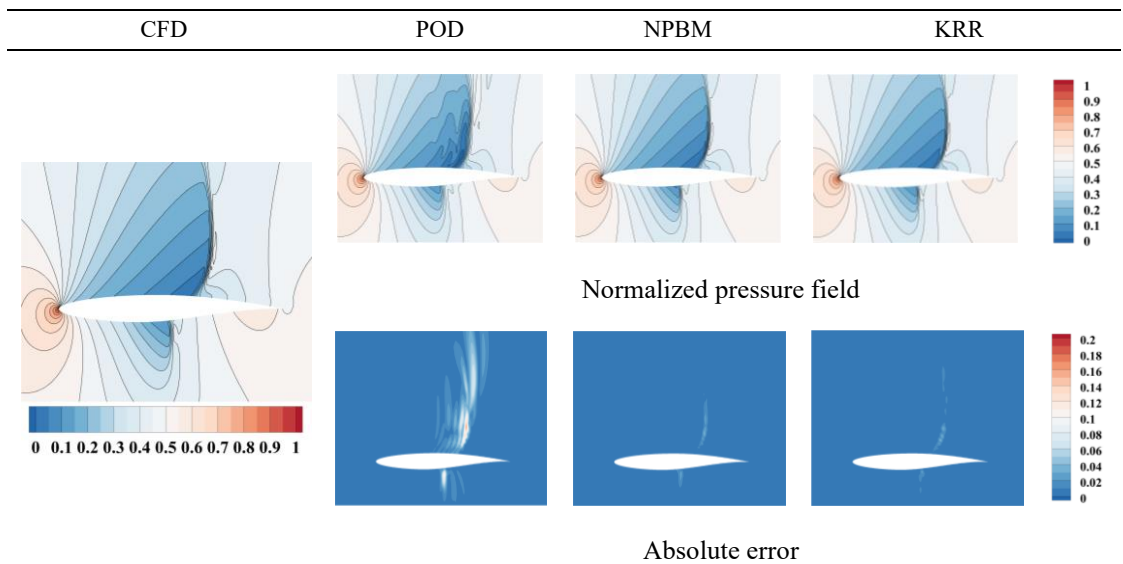

Figure 16 Comparison of the flow field reconstructed by different methods at $M_\infty = 0.804$, $\alpha = 2.05°$ and comparison with original flow field

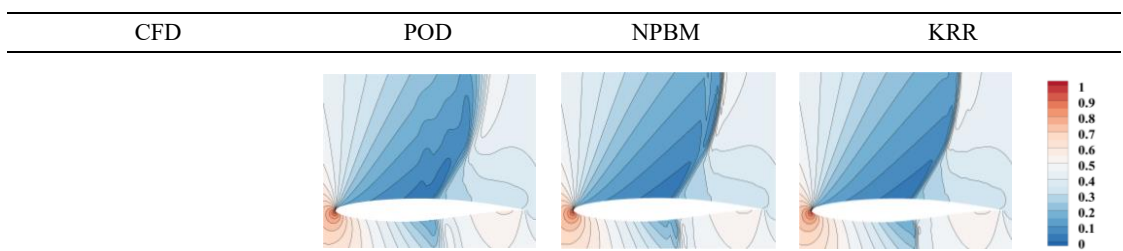

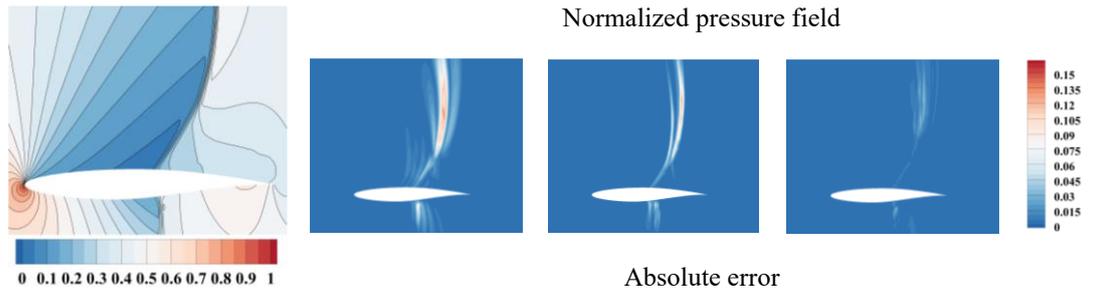

Figure 17 Comparison of the flow field reconstructed by different methods at $M_\infty = 0.827$, $\alpha = 5.10°$ and comparison with original flow field

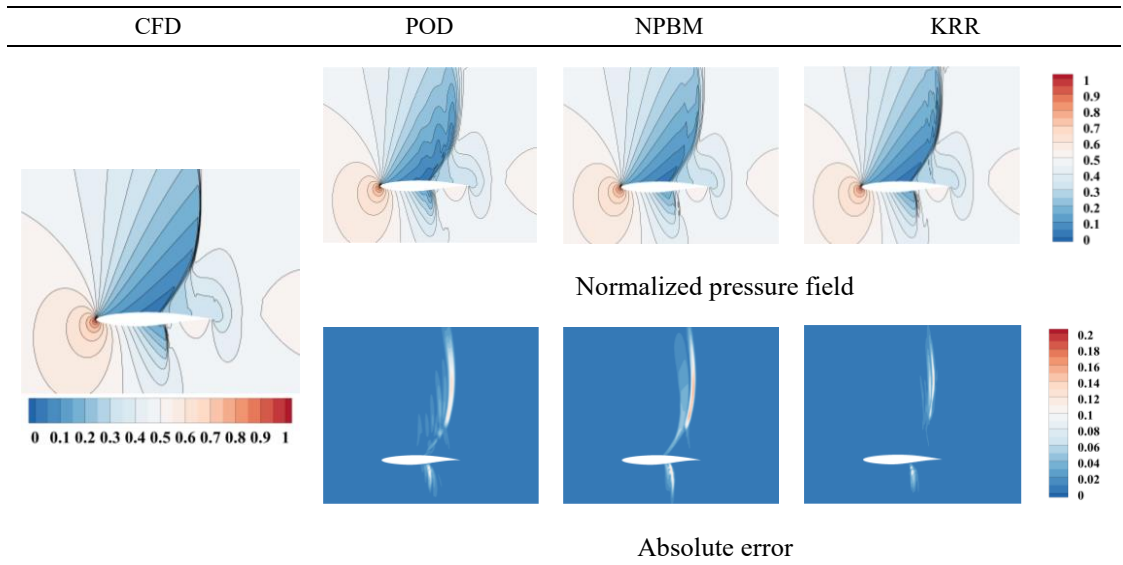

Figure 18 Comparison of the flow field reconstructed by different methods at $M_\infty = 0.843$, $\alpha = 5.55°$ and comparison with original flow field

Previously, we construct a set of nonlinear modes when applying KRR, then use these modes to reconstruct the flow fields. The first 9 modes of KRR are shown in Figure 19. The modes constructed by KRR contains rich local information around discontinuities, and each mode can capture the shock position and intensity of its corresponding training samples. The presence of these discontinuities' information means that the modes constructed by KRR is not smooth modes like that of POD, thereby enabling accurate reconstruction of the discontinuous regions and suppressing the oscillations in the reconstructed flow fields. However, when the model lacks sufficient discontinuities' information obtained from training samples in certain regions, it can lead to obvious oscillations in these regions, as discussed previously in Figure 18.

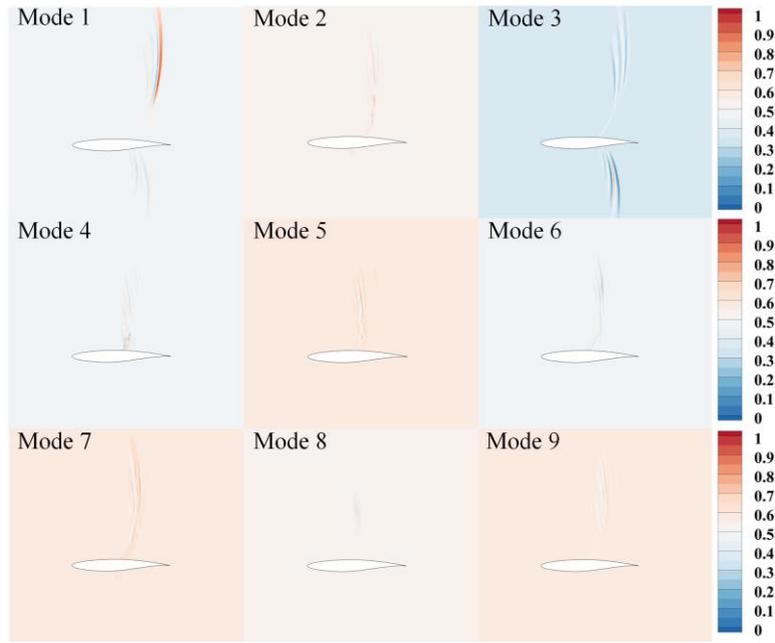

Figure 19 First 9 modes of KRR (normalized to (0, 1))

In comparison, the first 9 modes of POD are also shown in Figure 20. For POD, the information provided by modes 1-3 primarily focuses on the global characteristics of the flow field around the upper and lower surfaces, as well as the leading and trailing edges of the airfoil. As the order increases, the modes begin to gradually concentrate on the local discontinuities' information in the flow fields as shown in mode 4-9. However, since the discontinuous information is dispersed across multiple high-order modes, truncating the high-order modes may lead to the loss of a significant amount of localized discontinuities' information, resulting in larger reconstruction errors around shocks.

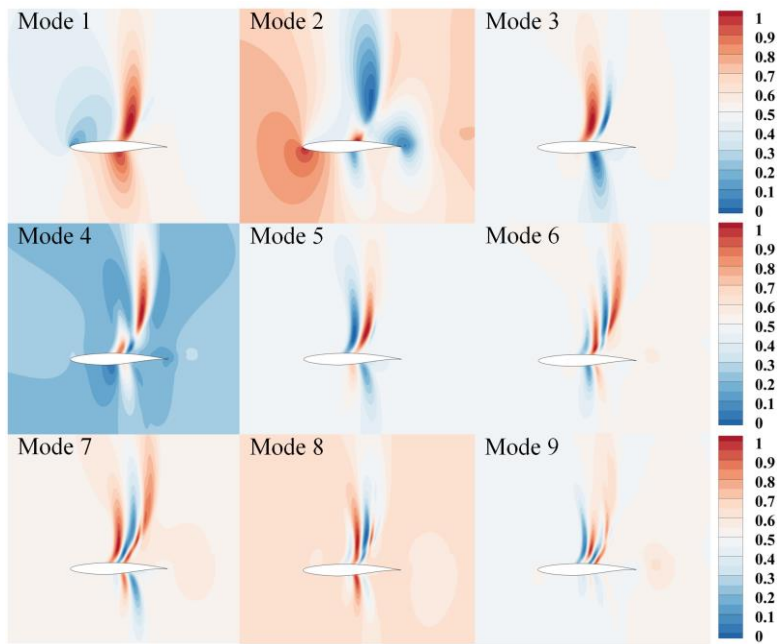

Figure 20 First 9 modes of POD (normalized to (0, 1))

# 4. Conclusion

In this paper, a new reconstruction method for flow fields with discontinuities is presented. The main conclusions are draw as follows:

1. The proposed method introduces KRR to construct a set of nonlinear modes, enabling accurate reconstruction of flow fields with discontinuities after dimensional reduction through nonlinear ROMs. These modes have strong physical interpretability, with each mode effectively capturing the discontinuities' information of its corresponding training sample. These local discontinuities' information allows the model to accurately reconstruct the flow field in the discontinuous regions.

2. The reconstruction accuracy of the flow fields using the proposed method is related to the dimensionality reduction method, the type of kernel function, and two hyperparameters $\lambda$ and $\gamma$. After comparison, we recommend using LLE to reduce the original flow fields to 17 dimensions and employing a quartic polynomial kernel function with $\lambda=10^{-5}$ and $\gamma=1.1$, which can achieve the minimum reconstruction error, to perform flow field reconstruction.

3. The proposed method significantly improves the reconstruction accuracy of the discontinuous regions in the wall pressure and the entire flow field for the RAE2822 airfoil at transonic. When at high $M_\infty$ and large $\alpha$, the proposed method can also accurately capture the shock's position and reconstruct the discontinuous regions.

4. On going work includes improving the reconstruction accuracy of the proposed method when the training samples are limited, and exploring the application of this method in predicting transonic and supersonic flow fields

# 5. Acknowledgements

The research was supported by the National Natural Science Foundation of China (NSFC Grant No. U2141254). The authors thankfully acknowledge the major advanced research project of Civil Aerospace from State Administration of Science Technology and Industry of China.

## Appendix A. The main process of ISOMAP and LLE

ISOMAP and LLE are two of the most representative manifold learning (ML) methods. Although both methods aim to discover the low-dimensional manifold structure of high-dimensional data, they differ significantly in approaches. According to the classification in reference [37], ISOMAP is a distance-preserving method that aims to ensure that the distances between samples remain consistent in the manifold space and the original space. In contrast, LLE is a topology-preserving method that achieves dimensionality reduction by maintaining the linear relationship between a sample and its neighbors.

Assume that the $i$th high-dimensional sample is vectorized to a vector $x_i$, and all of these samples form a matrix $\mathbf{X} \in \mathrm{R}^{n \times m}$, where $n$ is the number of samples and $m$ is the dimension of $x_i$. The process of ISOMAP and LLE are introduced as follow.

ISOMAP originates from the multiple dimensional scaling (MDS) by replacing the Euclidean distance with geodesic distance, thereby capturing nonlinear manifolds. ISOMAP consists of three steps: first, find the $k$ nearest neighbors of each sample $x_i$ based on the Euclidean distances, and based on these neighborhoods, a neighboring graph **G** is constructed. In **G**, the $i$th and $j$th sample*s* are connected by an edge with a weight of $d(i, j)$, where $d$ refers to the Euclidean distance between these two samples. Second, the shortest distance algorithm (usually the Floyd-Warshall algorithm[38]) is used to calculate the shortest distances matrix $\mathbf{D_G}$ between all sample on **G**. Finally, obtain the low-dimensional manifold coordinates matrix $\mathbf{Y} \in \mathbf{R}^{n \times p}$ by MDS, where $p$ is the dimension of the obtained manifold. This is achieved by

$$\min_{\mathbf{Y}} \left\| \mathbf{YY}^\mathrm{T} - \mathbf{B} \right\|_\mathrm{F}^2 \tag{15}$$

where the Gram matrix $\mathbf{B} = -(1/2)\mathbf{H}\mathbf{D_G}^2\mathbf{H}$, with centering matrix $\mathbf{H} = \mathbf{I}_n - (1/n)\mathbf{1}\mathbf{1}^\mathrm{T}$, $\mathbf{I}_n \in \mathbf{R}^{n \times n}$ the identity matrix, **1** the all-ones vector, $\left\| \cdot \right\|_\mathrm{F}$ the Frobenius norm and $(\cdot)^2$ the element-wise product.

LLE assumes that each sample can be reconstructed as a linear combination of its neighbors, and this linear relationship can be preserved in the low-dimensional manifold space. LLE The implementation process of LLE also consists of three steps. First, find the $k$ nearest neighbors of each sample $x_i$ and obtain their index $Q_i$, just like in ISOMAP. Second, for each sample $x_i$, solve the weight $w_{ij}$ of its $j$th neighbor by

$$\min_{w_1, w_2, \ldots, w_n} \sum_{i=1}^{n} \left\| \boldsymbol{x}_i - \sum_{j \in Q_i} w_{ij} \boldsymbol{x}_j \right\|_2^2 \tag{16}$$
$$\text{s.t.} \quad \sum_{j \in Q_i} w_{ij} = 1$$

where $\left\| \cdot \right\|_2$ is the 2-norm of a vector. Finally, through the computed weight $w_{ij}$, compute the low-dimensional manifold coordinates $\boldsymbol{y}_i$ for each sample by

$$\min_{y_1, y_2, \ldots, y_n} \sum_{i=1}^{n} \left\| \boldsymbol{y}_i - \sum_{j \in Q_i} w_{ij} \boldsymbol{y}_j \right\|_2^2 \tag{17}$$

then the manifold coordinates of all samples form the matrix $\mathbf{Y} \in \mathbf{R}^{n \times p}$, where $p$ is the dimension of the obtained manifold.

## Appendix B. The accuracy verification of CFD solver

The comparison of wall pressure coefficient on RAE2822 airfoil between CFD and experimental data at $M_\infty = 0.734$, $\alpha = 2.79°$ and $Re = 6.5 \times 10^6$ is shown in Figure 21. The computational mesh and CFD setups are the same with section 3.1. The experimental data is obtained from reference [39]. A good agreement between the CFD and experiment is demonstrated.

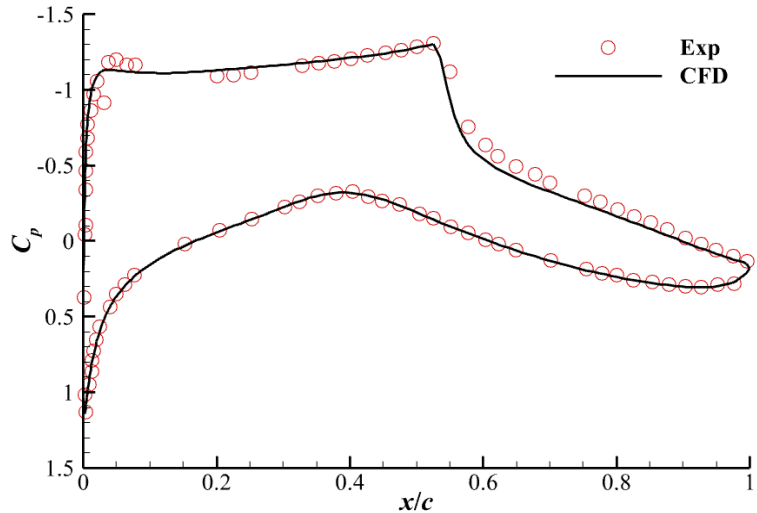

Figure 21 The comparison of wall $C_p$ between CFD and experiment